\documentclass[a4paper,12pt]{article}
\usepackage[utf8]{inputenc}
\usepackage[T1,T2A]{fontenc}
\usepackage{amsmath,amsfonts,amssymb}
\usepackage{cite}
\usepackage{graphicx}
\usepackage{wrapfig}

\begin{document}

\renewcommand*{\thefootnote}{\fnsymbol{footnote}}

\begin{center}
{\Large\bf Dynamical $\mathbf{O(4)}$-Symmetry in the Light Meson Spectrum within the Framework of the Regge Approach}
\end{center}
\bigskip

\begin{center}
{Sergey Afonin\(^{a,b}\)
and
Alisa Tsymbal\(^{a}\)
}
\end{center}

\begin{center}
  {\small\({}^a\)Saint Petersburg State University, 7/9 Universitetskaya nab.,
  St.Petersburg, 199034, Russia}\\
  \vspace*{0.15cm}
  {\small\({}^b\)National Research Center "Kurchatov Institute": Petersburg Nuclear Physics Institute,
  mkr. Orlova roshcha 1, Gatchina, 188300, Russia}
\end{center}

\renewcommand*{\thefootnote}{\arabic{footnote}}
\setcounter{footnote}{0}

\bigskip

\begin{abstract}
The light mesons tend to cluster near certain values of mass. As was noticed almost twenty years ago, the emergent degeneracy is of the same type as the dynamical $O(4)$-symmetry of the Coulomb potential in the hydrogen atom. The meson mass spectrum can be well approximated by the linear Regge trajectories of the kind $M^2=al+bn_r+c$, where $l$ and $n_r$ are angular momentum and radial quantum numbers and $a$, $b$, $c$ are parameters. Such a spectrum arises naturally within the hadron string models. Using 2024 data from the Particle Data Group, various fits for $M^2(l,n_r)$ were performed. Our analysis seems to confirm that $a\approx b$ in the light non-strange mesons, i.e., their masses depend on the sum $l+n_r$ as prescribed by the hydrogen-like $O(4)$-symmetry. Using the semiclassical approximation, we discuss on a simple qualitative level which kind of string-like semirelativistic approaches are more favored by the experimental data.
\end{abstract}

\bigskip

\section{Introduction}
The light hadron spectrum has been extensively studied since the appearance of the Quark Model more than 60 years ago. A large interest in this topic is not surprising --- the search for possible theoretical models describing the spectrum of the simplest bound states of quarks is tightly related with unveiling the nature of non-perturbative strong interactions and may lead to discoveries of new manifestations of strong interactions. The light meson spectrum is especially interesting in this quest since such particles consist of $u$- and $d$-quarks, just like nucleons.

The discrete spectrum of light meson resonances is somewhat simple: almost all known excited meson states included in the Particle Data Group (PDG) tables~\cite{pdg} are clustered near certain values of mass. Apparently, there are four pronounced clusters, see Table~\ref{tab1}.

\begin{table}
\caption{\small The approximate positions of clusters of light non-strange mesons according to relation~\eqref{sp} with the input parameters
$a=b=1.14~\text{GeV}^2$ and $c=0.5$, see the fit~\eqref{N}. Only the reliable states from Figure~1 were used (their amount for each cluster is indicated, the total amount is shown in brackets)
with unit weight.}
\label{tab1}
\begin{center}
\begin{tabular}{|c|c|c|}
\hline
\,\,$N$\,\,&\,\, $M$, MeV \,\,& \,\,Amount\,\,\\
\hline
0& $710$ & 2 (2) \\
1& $1280$ & 10 (12) \\
2& $1670$ & 8 (14)  \\
3& $1980$ & 21 (26) \\
4& $2250$ & 24 (30)  \\ \hline
\end{tabular}
\end{center}
\end{table}

The degeneracies in the spectrum of light meson excitations was broadly discussed about 15 years ago (see, e.g.,~\cite{klempt,cl1b,cl2,cl2b,cl2c,cl2d,cl3b,cl3c}). The following relation between the mass of a hadron and its quantum numbers can be written,
\begin{equation}
M^2=al+bn_r+c,
\label{sp}
\end{equation}
where $l$ and $n_r$ are the angular momentum and the radial quantum numbers, and $a$, $b$ and $c$ are parameters. It is important to notice that although the separation of total angular momentum $J$ into orbital angular momentum $L$ and intrinsic quark-antiquark spin $S$ is not Lorentz-invariant, the spin-orbital correlations seem to be highly suppressed inside excited light mesons because their typical lifetime is much less than the typical time required for noticeable effects of spin-orbital interactions. As a result, the standard quantum-mechanical relationship, $J=L+S$, is fulfilled with good accuracy. The $q\bar{q}$ pairs can form either singlet or triplet states with $S=0$ or $S=1$, respectively. However, it appears that the observed degeneracy is independent of $S$, and both singlet and triplet states are clustered together, with the positions of the clusters depending only on the angular momentum number $l$.

Long ago Chew and Frautschi observed that there exist linear Regge trajectories for angularly excited mesons. Later, it was observed that the daughter Regge trajectories are almost equidistant, i.e., the radial meson trajectories are approximately linear as well~\cite{ani}. The analysis of the experimental data showed that the slopes of angular and radial trajectories are close as it had been predicted by the Veneziano-like dual amplitudes and related hadron string models~\cite{cl2,cl2b}. In view of the appearance of new data in the last 15 years, it is interesting to reanalyze the experimental data and check to what extent they are compatible with the aforementioned exciting pattern of string-like spectrum.

The paper is organized as follows. The modern light meson spectrum is briefly overviewed in Section 2. In Section 3, we discuss a simple semiclassical hadron string picture explaining the
observed pattern of mass degeneracy. The last Section 4 contains a short summary of our results.

\section{The meson spectrum}

We will use the standard PDG tables~\cite{pdg} for our analysis. Since there is not enough well established data we will also include resonances from the "Further States" section of PDG.
Many mesons in that section were observed in the Crystal Barrel experiment and need further confirmation.

The internal quark structure of some meson resonances is not known yet. When constructing a classification of observed light mesons, a serious difficulty is the separation of states with
dominant strange component from the rest. The presence of $K$, $\bar{K}$, $\eta$ and $\phi$ mesons in the decay products does not mean that the strange component was dominant in the decayed meson: the final $s\bar{s}$-pair could be created from the vacuum triggering the strong decay, for instance, into $K\bar{K}$. Also the excited $\eta$-mesons may have a negligible
strange component due to the almost ideal mixing --- the label "$\eta$" may (and most likely) just denote the corresponding isosinglet. 
There are other difficulties related to the intrinsic structure of observed resonances --- a significant component of some of them might represent a tetraquark, meson molecule, or even glueball. A detailed discussion of each state will be given elsewhere. 

Some states with identical quantum numbers, such as $f_1(1285)$ and $f_1(1420)$ or $h_1(1170)$ and $h_1(1415)$, have relatively close masses (a typical excitation energy in hadrons is about 500~MeV). We will regard the heavier one as the state dominated by the $s\bar{s}$-component. 

The first task is to assign two quantum numbers, $n_r$ and $l$, to each observed state depending on its mass. Our assignments are shown in the Figure~1. Then, using the multiple linear regression method, several fits for Eq.(~\ref{sp}) were performed, with results shown in the Table~\ref{tab:fits}. Masses of resonances in the PDG tables are known with different precision. Thus, weights in the fit should be different for them, depending on the reliability of the points included in the analysis.
\begin{figure*}
\label{tab:clas}
\caption{\small A variant of "Mendeleev's table" for light mesons with (presumably) dominant $u,d$-quark component based on the $(L,n)$-classification of the present analysis. The missing states are marked with a question. The pale background is used for poorly known states and the states which may contain a non-negligible strange component.}
\vspace{-1cm}
\hspace{-1.4cm}
\includegraphics[width=17cm]{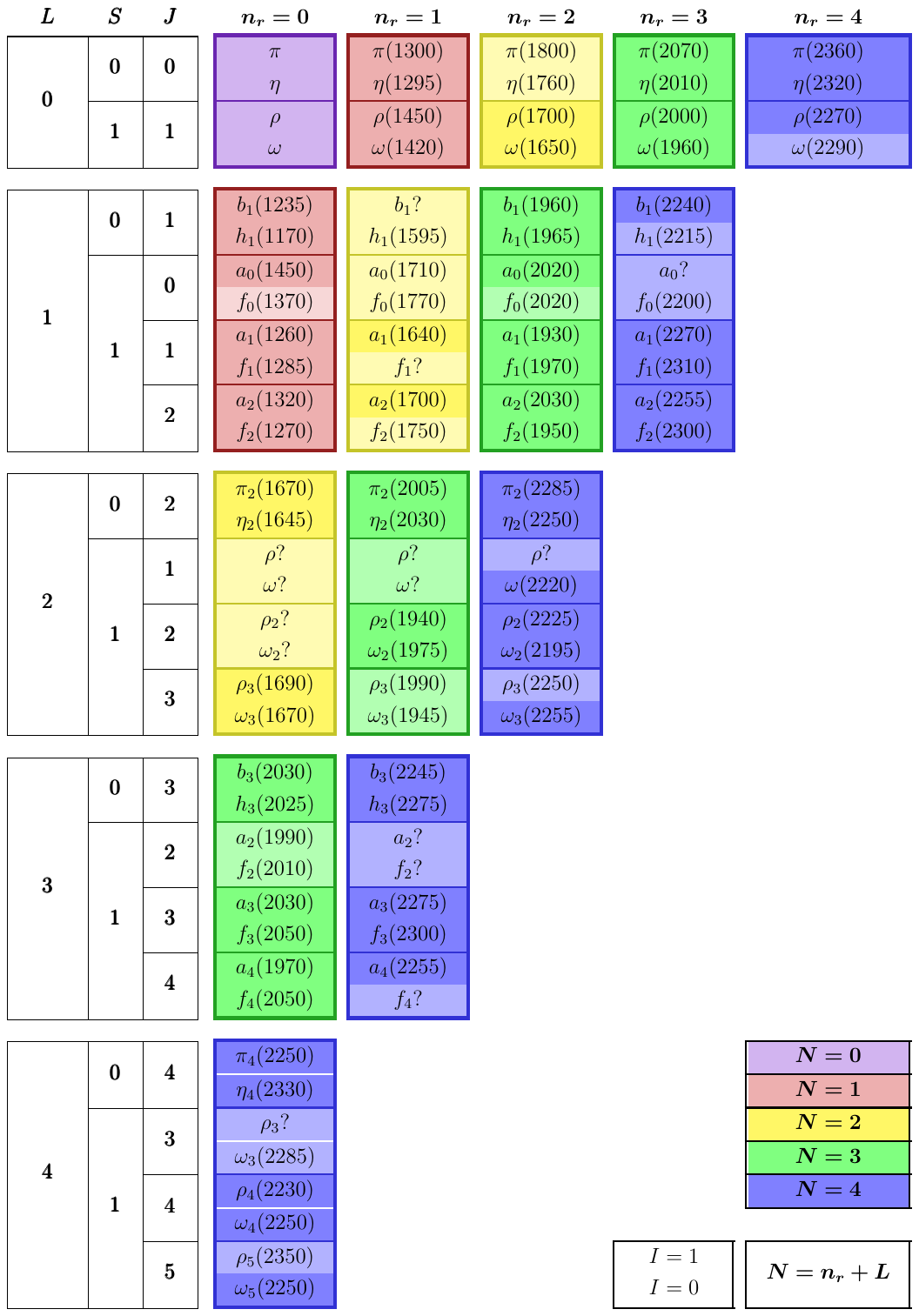}
\end{figure*}

We performed our analysis for two data sets. The first one includes only resonances with $I=1$, because they should not contain $s\bar{s}$ component by definition of isospin. The second data set contains all the resonances from the first data set and additionally all the mesons with $I=0$, which are believed to have relatively small $s\bar{s}$-component. The results obtained are slightly different for these two cases. In the second data set, the slope for the angular trajectory is a bit smaller than for the radial one. The difference, however, lies within the experimental uncertainty.

The first regression was done for the "raw" masses without any weighting. In the second and the third fit, the weight of the resonance is defined as $w_i=\frac{1}{\sigma_i^2}$. As the parameter $\sigma_i$ we took the experimental error $\Delta M^2_{\text{exp}}$ known from the PDG table in the second fit and, in the third one, the difference $\Delta M^2$ between the average mass squared in each cluster and the observed mass squared for each state. In the last fit, the weighting function was set to $\exp(-A\frac{\Delta M^2_{\text{exp}}}{M^2})$. The parameter
$A$ was chosen from the following conditions. Firstly, if the resonance is well-established and its mass is known with high precision, the weight must be close to $1$. Secondly, the weighting function should not fall off too rapidly. By changing the parameter $A$, one can obtain slightly different weight distributions, we used the case $A=7$ as a typical one.
The ground states corresponding to $n_r=l=0$ were excluded in all fits. The results of aforementioned fits are collected in the Table~2.

\begin{table}[h!]
\caption{\small Examples of fits for the relation $M^2=al+bn_r+c$.}
\label{tab:fits}
\bigskip
\begin{tabular}{|c|c|c|c|c|c|c|c|}
\hline
$I$ &\,\, Weight\,\, & $a$ & $b$ & $c$ &\,\, $\chi^2/\text{DoF}$\,\, \\ \hline
$1$ & none & $1.08\pm0.03$ & $1.13\pm0.03$ & \,\,$0.69\pm0.09$\,\, & 0.0052 \\
 & $\frac{1}{(\Delta M^2_{\text{exp}})^2}$ & $1.10\pm0.03$ & $1.16\pm0.03$ & $0.63\pm0.03$ & 0.0050 \\
 & $\frac{1}{(\Delta M^2)^2}$ & \,\,$1.113\pm0.005$\,\, &\,\, $1.126\pm0.005$\,\, & $0.62\pm0.01$ & 0.0051 \\
 & \,\,$\exp(-7\frac{\Delta M^2_{\text{exp}}}{M^2})$\,\, & $1.08\pm0.03$ & $1.13\pm0.03$ & $0.69\pm0.09$ & 0.0051 \\ \hline
$0$   & none & $1.13\pm0.20$ & $1.15\pm0.04$ & $0.58\pm0.03$ & 0.0045 \\
and & $\frac{1}{(\Delta M^2_{\text{exp}})^2}$ & $1.14\pm0.03$ & $1.13\pm0.04$ & $0.51\pm0.03$ & 0.0057 \\
$1$ & $\frac{1}{(\Delta M^2)^2}$ & $1.12\pm0.11$ & $1.16\pm0.09$ & $0.52\pm0.03$ & 0.0051 \\
 & $\exp(-7\frac{\Delta M^2_{\text{exp}}}{M^2})$ & $1.13\pm0.08$ & $1.15\pm0.03$ & $0.56\pm0.08$ & 0.0045 \\
  \hline
\end{tabular}
\end{table}

The obtained average slopes for $l$ and $n_r$ turned out to be very close, numerically their values are (in GeV$^2$),
\begin{align*}
&a=1.09\pm0.02,\quad b=1.14\pm0.02\qquad\text{ for the first data set, }\\
&a=1.13\pm0.05,\quad b=1.15\pm0.07\qquad\text{ for the second one.}
\end{align*}
Thus we can conclude that, at least in the first approximation, the whole spectrum seems to depend on a linear combination of $n_r+l$ as prescribed by the $O(4)$-symmetry. As in the hydrogen atom, we can introduce the principal quantum number $N$,
\begin{equation}
N\equiv l+n_r,\qquad N=0,1,2,\dots,
\end{equation}
and write the approximate spectrum as
\begin{equation}
\label{N}
M^2\approx1.14\, N+0.5,
\end{equation}
where $M^2$ is given in GeV$^2$. Strictly speaking, $N=n_r+l+1$ in the hydrogen atom, but here it is convenient to start $N$ with zero value.

\section{Hadron string models}
\par Various hadron string models are often exploited for the description of radially and angularly excited mesons. The idea is to consider a quark and antiquark as massless string ends bonded together with a gluonic flux tube. Then, the radial excitation of this string can be compared to the radially excited meson state, and the same may be done for the angular excitations.

Using hadron string models, the spectrum as in~Eq.(\ref{sp}) can be obtained from simple semiclassical considerations. In the reference frame in which the meson is at rest, the meson mass operator $\widehat{M}$ equals the hamiltonian $\widehat{H}$.
For two ultrarelativistic quarks with small masses, the Bethe-Salpeter equation may be written, which is a relativistic analog of the Schr\"{o}dinger equation,
\begin{equation*}
\left(\sqrt{\hat{p}^2+m_{q_1}^2}+\sqrt{\hat{p}^2+m_{\bar{q_2}}^2}+V(r)\right)\psi=E\psi.
\end{equation*}
The masses of $u$ and $d$ quarks are much smaller than their momenta, $\frac{m_q}{|p|}\ll 1$, and are considered equal, $m_{q_1}=m_{\bar{q_2}}$. The relativistic kinetic term is then
\begin{equation*}
\sqrt{p^2+m_q^2}+\sqrt{p^2+m_{\bar{q}}^2}\approx2p.
\end{equation*}
In the above relation, the momentum operator contains both radial and angular parts, $\hat{p}=\sqrt{-\frac{d^2}{dr^2}+\frac{l(l+1)}{r^2}}$. Most often, the potential part of the hamiltonian is set equal to the famous Cornell potential,
\begin{equation*}
V(r)=\sigma r -C_F\frac{\alpha_s}{r}+c.
\end{equation*}
The first term here is the linear confining potential which dominates at large distances between the quark and the antiquark. In further analysis, it will play the main role, since it coincides with the typical potential of the string with tension $\sigma$.
The second term describes the Coulomb-like interaction between quarks, which comes into play at short distances when quarks undergo a gluonic exchange.  $\alpha_s$ is the strong running coupling, and the fact that it is not a constant should entail some complications. However, in the further analysis, the Coulomb-like term will always be negligible compared to the linear one. The constant $C_F$ depends on the number of colors, $C_F\equiv(N_c^2-1)/2N_c$, that yields $C_F=4/3$ for $N_c=3$.
The last term in the Cornell potential is a negative constant $c$. The necessity to add $c$ is dictated by the nonrelativistic limit of the Bethe-Salpeter equation, and numerically it depends on the type of interacting quarks. Like with the Coulomb-like term, this constant will not play a role in the further analysis.

\subsection{The Radial Quantum Number}
 
The case of pure radial excitations corresponds to $l=0$. In this case, the momentum operator contains only the radial part $\hat{p}\equiv\hat{p}_r$. The potential part of the hamiltonian, as discussed previously, consists of three terms. At large distances, the linear term $\sigma r$ dominates, as if it were a classic string. Then, the meson mass is given by the relation
\begin{equation}
M=2p_r+\sigma r.
\label{m1}
\end{equation}
The constant $\sigma$ is the effective string tension. If the maximal flux tube length is $\ell$, then the total mass of the string can is $M=\sigma\cdot\ell$. 
In the semiclassical approximation, the quark wave function is written in the form
\begin{equation*}
\psi(r,t) \sim e^{iEt}e^{-i\oint p(r)dr}.
\label{wf}
\end{equation*}
The Bohr-Sommerfeld quantization condition states
\begin{equation}
\oint p(r)dr=2\int\limits_{0}^{\ell}p_r dr=\pi(n_r+\gamma).
\label{b-s-rad}
\end{equation}
We should draw attention to the factor $\pi$ in the r.h.s. of this condition --- it is not usual $2\pi$ (as in~\eqref{b-s-rad2} below) because the quark wave function must be antisymmetric,
consequently, the change of its phase after one period is equal to $\pi$.
Substituting the radial momentum from~\eqref{m1}, $p_r=\frac{M-\sigma r}{2}$, to the condition~\eqref{b-s-rad} one obtains the expected linear relation,
\begin{equation}
\label{m2}
M^2=2\pi\sigma(n_r+\gamma).
\end{equation} 
The constant $\gamma$ is of order of unity, and it depends on the turning points of the system, which are the ends of the string. In the case of central potentials $\gamma=\frac{1}{2}$.

The antisymmetry of the wave function mentioned above is very important~\cite{ar}: In most of the previous works (for example, in~\cite{cl2b,veseli,bicudo}), this was not taken into account, as a result the obtained slope between $M^2$ and $n_r$ was equal to $4\pi\sigma$ instead of $2\pi\sigma$.

There is another conceptual way to get the needed linear relation between $M^2$ and $n_r$. In this approach, the radial excitations are regarded as arising from collective gluon excitations the quarks are exchanging with (a kind of scalar "pomeron") at a constant separation --- a classical analogue of this is the vibrational excitation of a string with constant length. The kinetic term is now twice smaller (one quantizes the motion of only one effective particle), thus $p_r=M-\sigma r$. The quantization condition for a boson particle takes the form
\begin{equation}
\oint p(r)dr=2\int\limits_{0}^{\ell}p_r dr=2\pi(n+\gamma).
\label{b-s-rad2}
\end{equation}
Although this approach is conceptually different from the previous one, the result is the same --- the relation~\eqref{m2}.

\subsection{The Angular Momentum}

The spectrum of the rotating relativistic string is known to be well described by the Chew-Frautschi formula,
\begin{equation}
\label{m3}
M^2=2\pi\sigma l.
\end{equation}
We remind the reader the classical derivation of this relation.
Consider a gluonic flux tube as a solid body of known length $\ell$, rotating at the speed $v(r)=2r/\ell$. The total mass $M$ and the angular momentum $l$ of such a tube may be calculated as
\begin{equation*}
M=2\int\limits_0^{\ell/2}\frac{\sigma dr}{\sqrt{1-v^2(r)}}=\frac{\pi\sigma \ell}{2},\qquad l=2\int\limits_0^{\ell/2}\frac{\sigma rv(r) dr}{\sqrt{1-v^2(r)}}=\frac{\pi\sigma \ell^2}{8}.
\end{equation*}
Then, combining these two equations, one gets the Chew-Frautschi formula. This nice result was first derived by Nambu~\cite{nambu}. The next step is to impose the quantization
condition: $l$ in~\eqref{m3} must be integer, $l=0,1,2,\dots$. The logic here is the same as in the case of the hydrogen atom: One can first solve the corresponding classical problem and after that impose the quantization, the final result coincides with the one obtained within the framework of more accurate theory and the full procedure can be justified via the use of the semiclassical Bohr-Sommerfeld quantization. It is remarkable that the coefficient between $M^2$ and the angular momentum, $2\pi\sigma$, is the same as for the radial spectrum, hence, the final spectrum can be written in the following form,
\begin{equation}
\label{m4}
M^2=2\pi\sigma(n_r+l)+c.
\end{equation} 
It is interesting to mention that this spectrum (with certain constant $c$) follows from the strict quantization of Nambu-Goto string with tension $\sigma$.

It should be remarked that in the literature one often uses another semiclassical method for getting the linear relation between $M^2$ and the angular momentum:
the starting picture is the circular motion of two point quarks interacting via the linear potential at the constant distance $r$. The Bohr-Sommerfeld quantization condition~\eqref{b-s-rad2}
for such a circular motion of two particles then reads
\begin{equation}
2\oint p(r)dr=2\int\limits_{0}^{2\pi}p\,\frac{r}{2}\, d\phi=2\pi l,\qquad l=0,1,2,\dots.
\label{b-s-rad3}
\end{equation}
Note that the antisymmetric nature of quarks is taken into account: since the wave function of the system includes the product of two quark wave functions and each one acquires
the phase $\pi$ after one full turn, the product acquires the phase $2\pi$. The condition~\eqref{b-s-rad3} leads to the standard semiclassical quantization for the angular momentum,
$pr=l$. Substituting this condition and the second Newton's law,
\begin{equation*}
\frac{\partial V(r)}{\partial r}=\sigma=ma=m\frac{v^2}{r}=E_{\text{kin}}\frac{2}{r}=p\,\frac{2}{r}\,,
\end{equation*}
into $M^2=(2p+\sigma r)^2$ one gets~\cite{bicudo}
\begin{equation}
M^2=8\sigma l.
\label{m5}
\end{equation}
The given derivation of angular spectrum~\eqref{m5} is heuristic but a more accurate (and much more lengthy) analysis can be done which shows the validity of~\eqref{m5} for $l\gg1$.

The slope in~\eqref{m5} is different from the slope of string spectrum~\eqref{m4}. Thus the relation~\eqref{m5} contradicts the available experimental data. A physical reason
can be easily foreseen: In reality, the typical lifetime of meson excitations is too short for a quark-antiquark pair to perform a complete "turn" around a common center of mass. This
makes the second model unrealistic. On the other side, when quarks move past each other in hadron collisions, the forming gluon string between them must actually "turn" through a certain angle as it happens within the Nambu string picture. The latter yields the spectrum~\eqref{m3} even for infinitesimal rotation. For this reason the string-like approach is expected to better capture the underlying physics of resonance formation at non-zero quark angular momentum than the semirelativistic potential approach.

\section{Conclusions}

Our new Regge analysis of modern experimental data on the light mesons confirms the existence of a broad mass degeneracy in the light non-strange mesons.

In the first approximation, this spectrum depends linearly on the radial quantum number $n_r$ and the orbital number $l$, with
\begin{equation*}
M^2=al+bn_r+c.
\end{equation*}

Our analysis of the experimental data confirms the approximate universality of the angular and radial slopes with the averaged common value of $a\approx b\approx 1.14$~GeV$^2$. This means that the considered meson mass spectrum must depend on a single quantum number $N=n_r+l$. This leads to the $O(4)$-like degeneracy of the same kind as in the hydrogen atom.

The observed degeneracy has a simple qualitative explanation within the framework of semiclassical hadron string approach resulting in the relation $a=b=2\pi\sigma$ for the frequencies of orbital and oscillatory motions.

\section*{Acknowledgments}

The useful discussions with A. Sarantsev are gratefully acknowledged.

\end{document}